\documentclass{article}
\usepackage[utf8]{inputenc}
\usepackage{amsmath,amsthm, amsfonts,bm,graphicx,url}
\usepackage{color, colortbl}

\newcommand{\ep}{\varepsilon}

\newcommand{\Bi}{B^{\text{init}}}
\newcommand{\Be}{B^{\text{end}}}

\newtheorem{theorem}{Theorem}[section]

\newcommand{\brac}[1]{\left(#1\right)}
\title{Failure of monotonicity in epidemic models}
\author{Maria Chikina and Wesley Pegden}
\date{April 2020}

\begin{document}

\maketitle

\begin{abstract}
    We discuss the failure of monotonicity properties for even simple compartmental epidemic models, for the case where transmission rates are non-constant. We also identify a special case in which monotonicity holds.
\end{abstract}

\section{Introduction}
Consider a simple SIR (Susceptible, Infected, and Recovered/Removed) model governed by the differential equations

\begin{align}
\label{dIdt}\frac{dI}{dt}&=I\beta S\cdot \frac{1}{m}-\alpha I\\
\label{dSdt}\frac{dS}{dt}&=-I\beta S\cdot \frac{1}{m} \\
\label{dRdt}\frac{dR}{dt}&=\alpha I.\\
\end{align}

When the transmission coefficient $\beta$ and the recovery coefficient $\alpha$ are constant (as is the total population $m$), the final recovered population $R(\infty):=\lim_{t\to \infty} R(t)$ (which is the total number of people ever infected) can be found to satisfy the transcendental equation
\[
R(\infty)=1-S(0)e^{-\tfrac \beta {\alpha m}\brac{R(\infty)-R(0)}},
\]
by using the chain rule to write
\[
\frac{dS}{dR}=-\frac{\beta S}{\alpha m},
\]
separating the variables, and integrating.  From this it follows that $R(\infty)$ behaves monotonically with respect to the basic reproduction number $R_0:=\frac{\beta}{\alpha}$; that is, the larger $R_0$, the larger $R(\infty)$.

The purpose of this note is consider monotonicity properties obeyed by the system in a context where the transmission rate is not a constant but also a function of time (because of effects of seasonality, for example, or the effects of mitigations).  We will see that natural-seeming monotonicity properties fail to hold for even this simplest of epidemic models.  


For an SIR model where the transmission rate $\beta:[0,\infty]\to \mathbb{R}$ is a function of time,
consider the following three monotonicity properties one might hope would be satisfied for such a system, if $\beta_1$ and $\beta_2$ are two transmission rate functions, and $R_1(\infty)$ and $R_2(\infty)$ are the recovered populations resulting from SIR models with $\beta_1$ and $\beta_2$, respectively.\\

\noindent \textbf{Pointwise monotonicity}: If $\beta_1(t)\leq \beta_2(t)$ for all $t$, then $R_1(\infty)\leq R_2(\infty)$.\\

\noindent \textbf{Shift monotonicity}: If $\beta_1(t)=\beta_2(t)-\ep$ for all $t$ for some constant $\ep>0$, then $R_1(\infty)\leq R_2(\infty)$.\\

\noindent \textbf{Scaling monotonicity}: If $\beta_1(t)=(1-\ep)\beta_2(t)$ for all $t$ for some constant $\ep>0$, then $R_1(\infty)\leq R_2(\infty)$.


\bigskip

In this note we show that \textbf{all three of these forms of monotonicity can fail}, in general, for SIR models where the transmission rate $\beta(t)$ is not constant. 

On the other hand, we show that pointwise monotonicity \emph{does} hold, in the special case that transmission rates never increase:

\begin{theorem}\label{t.mono}
Consider functions $\beta_1,\beta_2$ which are both bounded below by some positive constant.  If $\beta_1(t)\leq \beta_2(t)$ for all $t$ and $\beta_1(t)$ is monotone nonincreasing, then an SIR model satisfying \eqref{dIdt}, \eqref{dSdt}, \eqref{dRdt}, governed by the transmission rate function $\beta:=\beta_1$ ends with a greater number of susceptible individuals than when governed by $\beta:=\beta_2$ (assuming identical initial conditions).
\end{theorem}

We remark that the lack of monotonicity we point out in this note could be more severe in more complex, realistic epidemic models, where, for example, the presence of heterogeneous transmission and mortality rates can lead to substantial benefits to increasing transmission rates for some populations \cite{age}.


\section{Failures of monotonicity}
In the following sections we present examples to demonstrate the failure of each type of monotonicity.  The examples are generated using code available at the entry for this manuscript at \url{http://math.cmu.edu/~wes/pub.html}. 
Note that we have not tried to optimize these examples to maximize non-monotone effects; it is easy to find more egregious failures of monotonicity.

\subsection{Failure of pointwise monotonicity}
\begin{figure}
    \centering
    \includegraphics[width=.7\linewidth]{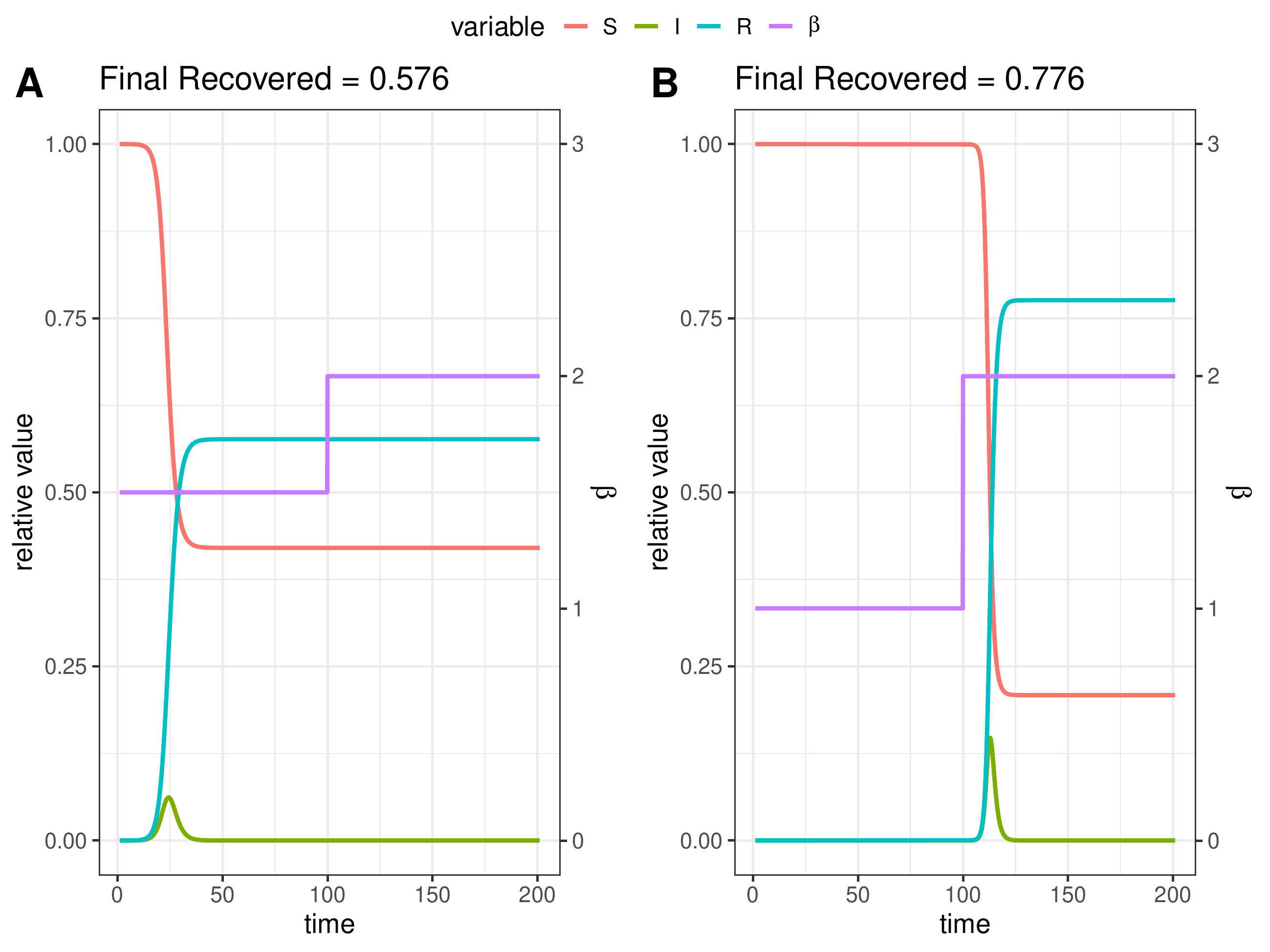}
    \caption{Failure of pointwise monotonicity.  The scale for $S$, $I$, and $R$ is on the left, and for $\beta$, on the right.}
    \label{fig:point}
\end{figure}

The failure of pointwise monotonicity for SIR models is well-known; see \cite{overshoot} for a recent reference discussing examples.  A standard example can be constructed by taking

\[
\beta_1(t):=\begin{cases}
1& t<100\\
2& t\geq 100
\end{cases},
\]
while
\[
\beta_2(t):=\begin{cases}
1.5& t<100\\
2& t\geq 100
\end{cases}.
\]
For convenience we scale time so that $\alpha=1$, here and for the remaining examples, giving the correspondence $\beta=R_0$.  The result is shown in Figure \ref{fig:point}.

Qualitatively, this particular example is typically explained qualitatively by noting the fact that $\beta_2(t)>\beta_1(t)$ for $t\in [0,100]$ results in an initial epidemic in the second scenario, which, while smaller than the epidemic for $\beta=2$ seen in the first scenario, is nevertheless large enough to achieve enough population immunity to resist transmission when $\beta_2(t)=2$.

However, it is important to note that there is no general principle asserting that this simple example is the only way in which monotonicity can fail; instead, this simple examples underlines the fact that pointwise monotonicity is not guaranteed by even the simplest forces at play in an epidemic's growth.  As we will see, monotonicity can fail in ways which are more surprising than this first example might suggest.
\subsection{Failure of shift monotonicity}
Perhaps more surprising than the failure of pointwise monotonicity is the fact that even consistent changes to the transmission function over time can be counterproductive for reducing epidemic size.  

\begin{figure}
    \centering
    \includegraphics[width=.7\linewidth]{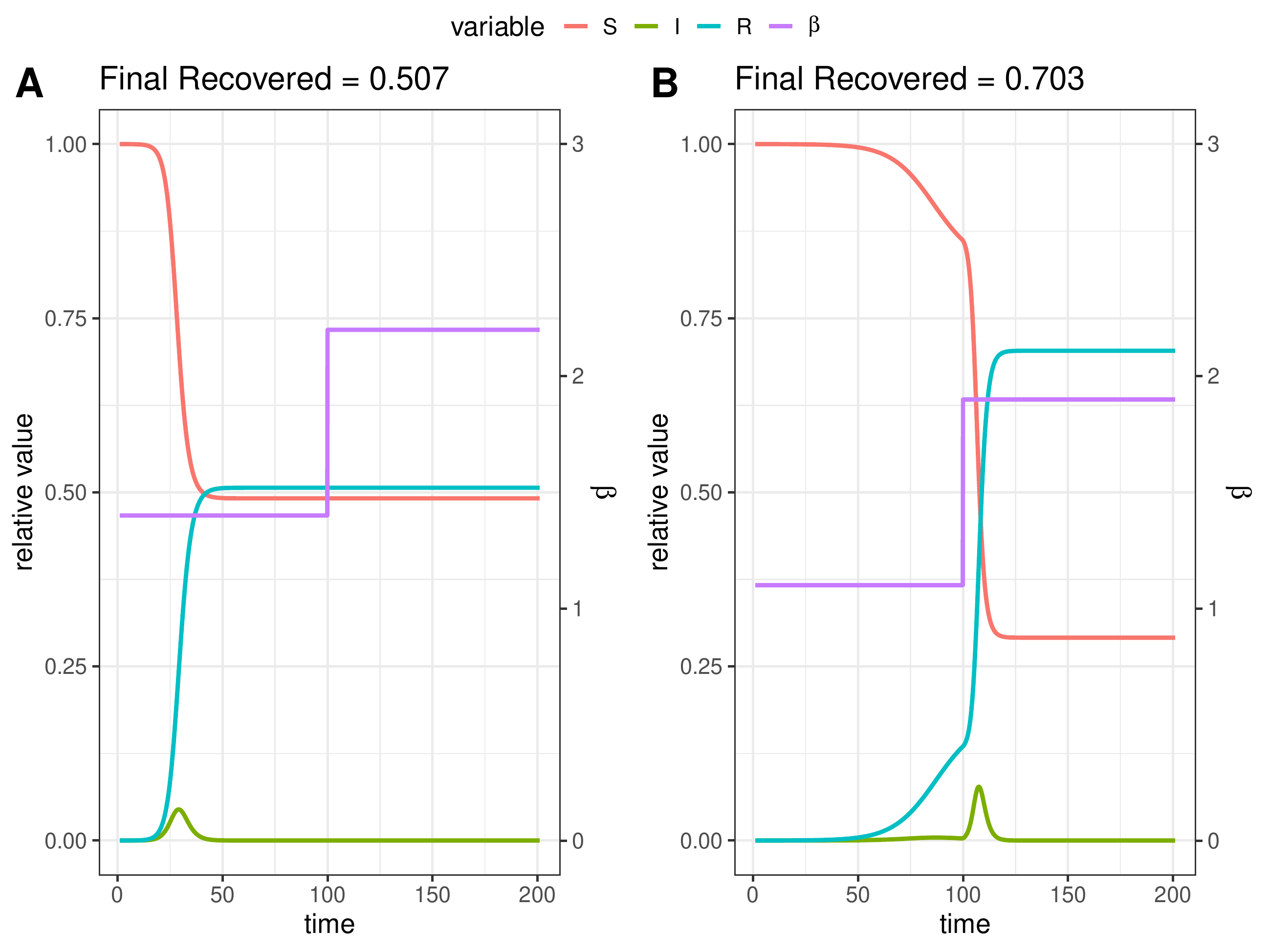}
    \caption{Failure of shift monotonicity.  The scale for $S$, $I$, and $R$ is on the left, and for $\beta$, on the right.}
    \label{fig:shift}
\end{figure}
Consider epidemics governed by the transmission functions
\[
\beta_1(t):=\begin{cases}
1.1& t<100\\
1.9& t\geq 100
\end{cases},
\]
while
\[
\beta_2(t):=\begin{cases}
1.4& t<100\\
2.2& t\geq 100
\end{cases}.
\]

Observe that $\beta_1(t)=\beta_2(t)-0.3$ for all $t$.   (Recall we have set $\alpha=1$.) Nevertheless, nearly 40\% more total infections occur in an epidemic governed by $\beta_1$ instead of $\beta_2$, as we see in Figure \ref{fig:shift}.  As we will see next, the same surprising result can occur from a geometric scaling, rather than an arithmetic shift.

\subsection{Failure of scaling monotonicity}
\begin{figure}
    \centering
    \includegraphics[width=.7\linewidth]{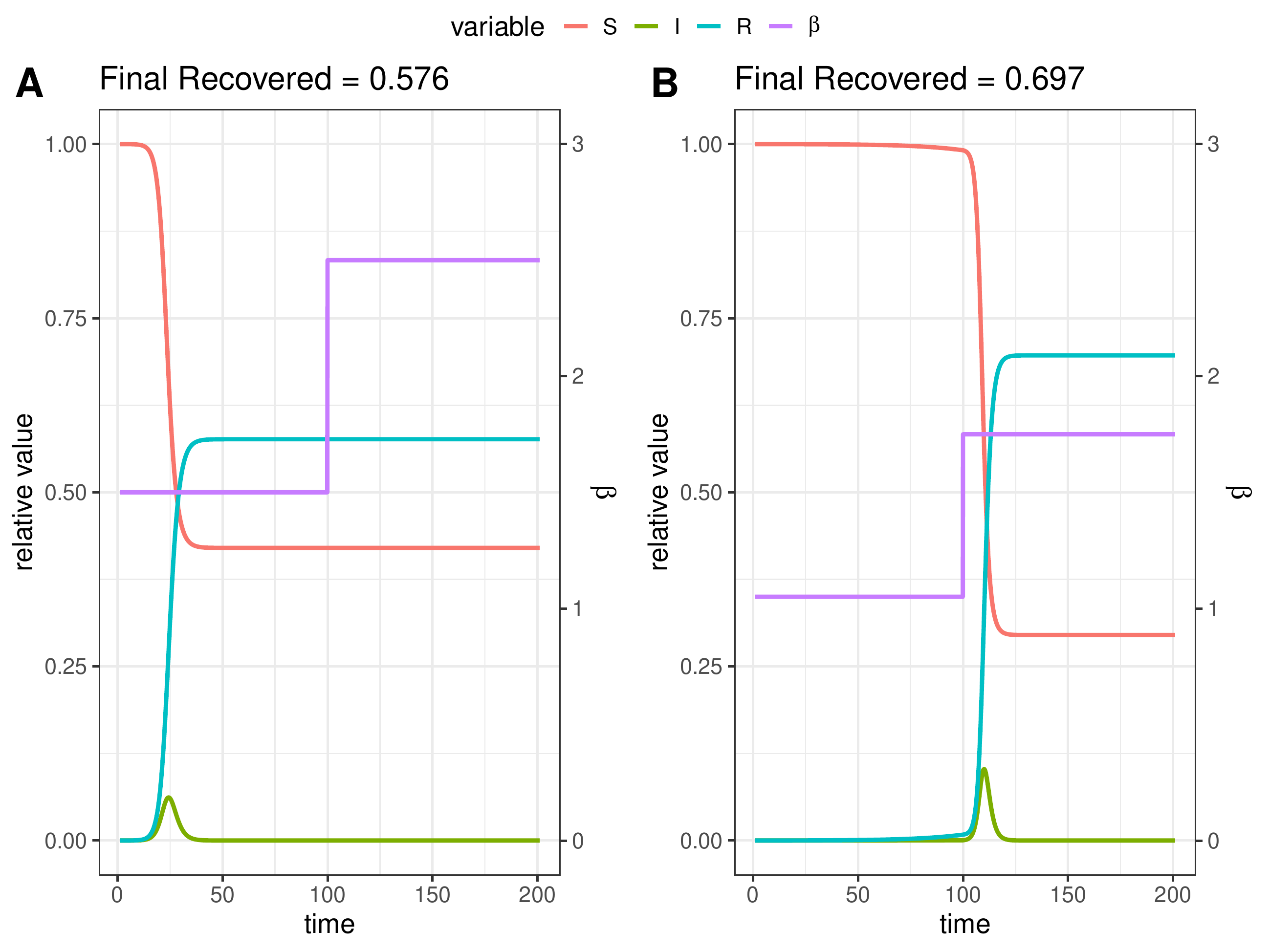}
    \caption{Failure of scale monotonicity.  The scale for $S$, $I$, and $R$ is on the left, and for $\beta$, on the right.}
    \label{fig:scale}
\end{figure}

Here we see the same type of example but where the global change in transmission rates is geometric.  

We let
\[
\beta_2(t):=\begin{cases}
1.5& t<100\\
2.5& t\geq 100
\end{cases}.
\]
and $\beta_1=.7\cdot \beta_2.$  The result is in Figure \ref{fig:scale}.

\subsection{Failure of shift for naturalistic transmission variation}

\begin{figure}
    \centering
    \includegraphics[width=.7\linewidth]{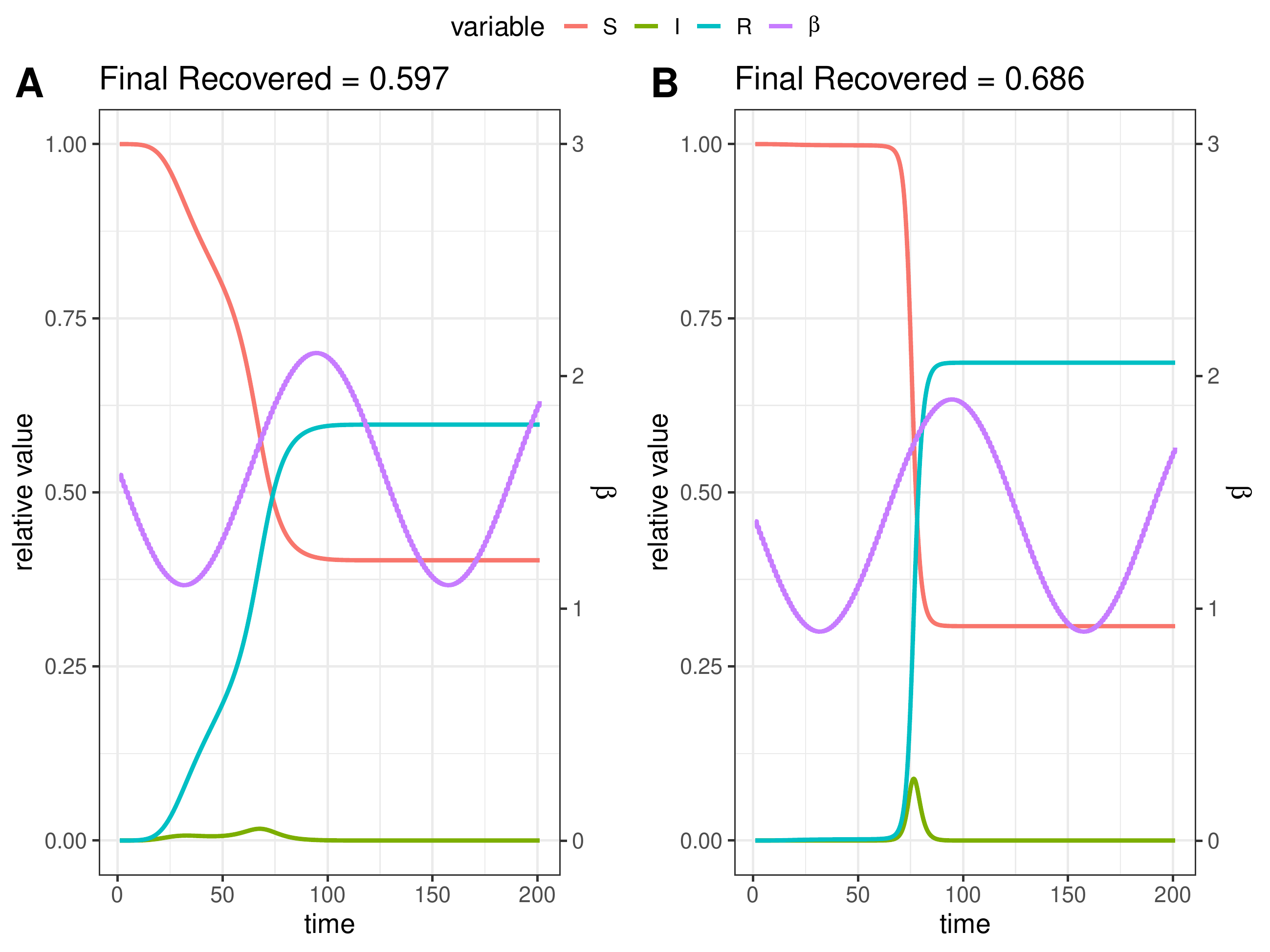}
    \caption{Failure of shift monotonicity for naturalistic variation. The scale for $S$, $I$, and $R$ is on the left, and for $\beta$, on the right.}
    \label{fig:sine}
\end{figure}

To produce examples which are as simple as possible, we have focused on step-functions in the previous sections.  This may give the incorrect impression that failure of monotonicity only occurs in the special case that there will be some sudden resumption of high transmission rates in the future.  In fact it is easy to construct examples where monotonicity fails even when transmission rates are changing only slowly over time.   In this section, we give an example motivated by the possibility of seasonal effects on viral transmission, using a shift of a sine-wave function as the time-varying transmission rate $\beta$.

We use
\[
\beta_2(t):=
\frac{\sin{t/20}+1} 2 +1.9
\]
and
\[
\beta_2(t):=
\frac{\sin{t/20}+1} 2 +2.1.
\]

Again, we see that a small consistent \emph{decrease} in transmission rates in this scenario can result in more total infections (Figure \ref{fig:sine}).

\section{Monotonicity when $\beta$ is decreasing}

In this section we prove Theorem \ref{t.mono}.   For this purpose, let the functions $S_i,I_i,R_i$ of $t$ for $i=1,2$ be the two SIR models satisfying the systems corresponding to \eqref{dIdt}, \eqref{dSdt}, \eqref{dRdt} for the given $\beta_i$.  Note that for simplicity Theorem \ref{t.mono} assumes that the differential equalities hold everywhere, which means that the $\beta_i$ are necessarily continuous.  For simplicity of notation we let $m=1$ without loss of generality, so that $I$, $S$, and $R$ are all ratios between 0 and 1.

As is standard, we can view these models under derivatives with respect to $S$ instead of $t$:  Beginning from \eqref{dIdt}, \eqref{dSdt}, \eqref{dRdt}, we can apply to the chain rule to get
\begin{equation}\label{dIdS}
\frac{dI}{dS}=\frac{\alpha}{\beta \cdot S}-1.
\end{equation}
We write $B=1-S$ to write 
\begin{equation}\label{dIdB}
\frac{dI}{dB}=1-\frac{\alpha}{\beta \cdot (1-B)}.
\end{equation}

Considering $I$ and $\beta$ as a functions of $B$, and assuming for simplicity that the number of initially recovered individuals is 0, the choice of initial value for $I$ is made by choosing some $\Bi<1$ such that $I(\Bi)=\Bi$.  (In the case where the initial number $r_0$ of recovered individuals is nonzero, set $I(\Bi)=\Bi-r_0>0$.)


Note that we already see from \eqref{dIdS} that the SIR model \emph{does} satisfy pointwise monotonicity with respect to $\beta$ \emph{as a function of $B$}; the spectacular failure of monotonicity for SIR models occurs precisely because of the interaction between $\beta$ and the correspondence between $S$ and $t$.

To proceed, we consider the derivative
\begin{equation}\label{dtdS}
\frac{dt}{dB}=\frac{1}{I\cdot \beta\cdot (1-B)}.
\end{equation}
Now we can define $f:\mathbb{R}^3\to \mathbb{R}^2$ by
\begin{equation}
    \label{e.myf} f(I,t,B)=\brac{1-\frac{\alpha}{\beta_1(t)\cdot (1-B)},\frac{1}{I\cdot \beta_1(t)\cdot (1-B)}}
\end{equation}
so that the SIR model governed by $\beta_1$ satisfies
\[
  \frac{d}{dB}(I_1,t_1)=f(I_1,t_1,B)\label{I2ode}.
\]
Observe that $f$ is Lipshitz in $[\Bi,\Be]$ since $B$ is bounded away from 1, and $\beta(t)$ and $I$ are bounded away from 0 on this range.

We write $f_I$ and $f_t$ for the first and second coordinates of $f$, respectively.   Observe that $f_I$ is decreasing with respect to $t$, and $f_t$ is decreasing with respect to $I$.  In particular, 
\[
g(I,t,B):=\bigg(-f_I(-I,t,B),\:f_t(-I,t,B)\bigg)
\]
is \emph{quasimonotone} in $(I,t)$; that is, $g_I$ is increasing in $t$ and $g_t$ is increasing in $I$.  

Moreover, we have that the SIR models governed by $\beta_1$ and $\beta_2$, respectively, satisfy
\begin{align}
    \frac{d}{dB}(-I_1,t_1)&= g(-I_1,t_1,B)\label{I1gode}\\
    \frac{d}{dB}(-I_2,t_2)&\leq g(-I_2,t_2,B),\label{I2gode}
\end{align}
where the inequality follows by our assumption that $\beta_1(t)\leq \beta_2(t).$

As a result, from the relations \eqref{I1gode}, \eqref{I2gode}, together with $I_1(\Bi)\leq I_2(\Bi),$ $t_1(\Bi)\geq t_2(\Bi)$ and the Lipshitz condition for $f$ (and so $g$) imply that
\begin{align*}
I_1\leq I_2\\
t_1\geq t_2
\end{align*}
throughout; see \cite{walter}, page 112, in particular version (b) of the \emph{Comparison Theorem} on the same page.\qed

\end{document}